\documentclass{article}
\usepackage{wrapfig}
\usepackage[nonatbib, final]{neurips_2022_ml4ps}

\usepackage[utf8]{inputenc} % allow utf-8 input
\usepackage[T1]{fontenc}    % use 8-bit T1 fonts
\usepackage{hyperref}       % hyperlinks
\usepackage{url}            % simple URL typesetting
\usepackage{booktabs}       % professional-quality tables
\usepackage{amsfonts}       % blackboard math symbols
\usepackage{nicefrac}       % compact symbols for 1/2, etc.
\usepackage{microtype}      % microtypography
\usepackage{xcolor}         % colors

\usepackage[numbers]{natbib}
\usepackage{amsmath} 
\usepackage{graphicx}
\usepackage{subcaption}
\usepackage{siunitx}
\usepackage{bigdelim} % rdelim
\usepackage{changepage}

\def\gfdl{{GFDL-ESM2G~}}

\newcommand{\yes}[0]{{\textcolor{blue}{ [Yes]~ }}}
\newcommand{\na}[0]{{\textcolor{gray}{ [N/A]~ }}}
\newcommand{\no}[0]{{\textcolor{orange}{ [No]~ }}}

\setcitestyle{numbers}

\title{Multi-scale Digital Twin: Developing a fast and physics-informed surrogate model for groundwater contamination with uncertain climate models}

\author{
    Lijing Wang \\
    Dept. of the Geological Sciences\\
    Stanford University\\
    Stanford, CA 94305 \\
    \texttt{lijing52@stanford.edu} \\
\And
        Takuya Kurihana \\
    Dept. of Computer Science\\
    University of Chicago\\
    Chicago, IL 60637 \\
    \texttt{tkurihana@uchicago.edu} \\
\And
    Aurelien Meray \\
    Dept. of Computer Science\\
    Florida International University\\
    Miami, FL 33133 \\
    \texttt{amera009@fiu.edu} \\
\And
    Ilijana Mastilovic \\
    School of Freshwater Sciences \\
    University of Wisconsin - Milwaukee \\
    Milwaukee, WI 53204 \\
    \texttt{ilijana@uwm.edu} \\
\And
    Satyarth Praveen \\
    Lawrence Berkeley National Lab \\
    Berkeley, CA \\
    \texttt{satyarth@lbl.gov} \\
\And
    Zexuan Xu \\
    Lawrence Berkeley National Lab \\
    Berkeley, CA \\
    \texttt{zexuanxu@lbl.gov} \\
\And
    Milad Memarzadeh \\
    NASA Ames Research Center \\
    Mountain View, CA \\
    \texttt{milad.memarzadeh@nasa.gov} \\
\And
    Alexander Lavin \\
    Pasteur Labs \& ISI \\
    Brooklyn, NY 11205 \\
    \texttt{lavin@simulation.science} \\
\And
    Haruko Wainwright \\
    Department of Nuclear Science and Engineering \\
    Massachusetts Institute of Technology \\
    Cambridge, MA \\
    \texttt{hmwainw@mit.edu} \\
}

\begin{document}
\maketitle
\begin{abstract}
Soil and groundwater contamination is a pervasive problem across all societies worldwide. 
Contaminated sites often require decades to remediate or to monitor natural attenuation. 
Climate change exacerbates the problem because extreme precipitation and/or shifts in precipitation/evapotranspiration regimes can remobilize contaminants and proliferate affected groundwater.
To quickly evaluate the spatiotemporal variations of groundwater contamination under uncertain climate disturbances, we developed a physics-informed machine learning surrogate model using U-Net enhanced Fourier Neural Operator (U-FNO) to solve Partial Differential Equations (PDEs) for complex flow and transport physical simulations for local contamination sites.

We develop a combined loss function that includes both data-driven factors and physical boundary constraints at multiple spatiotemporal scales. Our U-FNOs can reliably predict the spatiotemporal variations of groundwater flow and contaminant transport properties from 1954 to 2100 with realistic climate projections. 
In parallel, 
we develop a convolutional autoencoder combined with online clustering to reduce the dimensionality of the vast historical and projected climate data by quantifying climatic region similarities across the United States. 
The ML-based unique climate clusters provide realistic climate projections for the surrogate modeling and help return reliable future recharge rate projections immediately without querying large climate datasets. 
In all, this Multi-scale Digital Twin work can advance the field of environmental remediation under climate change.
\paragraph{}
\end{abstract}

\section{Introduction}\label{sec:introduction}
Contaminated sites are highly heterogeneous, varying in size and distribution, and contaminant types from industrial or agricultural activities, to nuclear and battery waste storage. 
They must be carefully managed to prevent hazardous materials from causing harm to humans, wildlife, and ecological systems. 
Additional challenges arise and exacerbate risks due to global climate change. Uncertain climate scenarios can impact the water balance in dynamic or disruptive ways and remobilize contaminants. In this paper, we look into advances in machine learning (ML) and physical simulation to predict the spatiotemporal contamination under future climate changes for decision-makers and site managers. We incorporate multi-scale uncertainties in physical simulations: uncertain global climate projections and uncertain local soil and subsurface properties. We run groundwater flow and contaminant transport physics simulations using Amanzi~\citep{moulton2014amanzi} at the testbed: the Department of Energy's Savannah River Site F-Area \citep{libera2019climate,xu2022reactive} with uncertain subsurface, soil and climate factors. However, solving a complex groundwater flow and contaminant transport model such as Amanzi takes long time to run even with supercomputers. Meanwhile, the global climate models \citep{knutti2013robustness} are in a different spatial scale than the local testbed.

Therefore, we tackle this multi-scale digital twin problem with two ML solutions: 1) \textbf{ML-based surrogate model}: we develop a ML surrogate model using neural operator learning \citep{li2020fourier,lavin21,wen2022u} to solve Partial Differential Equations (PDEs) for complex flow and transport physical simulations rapidly, 2) \textbf{Unsupervised climate data clustering}: we perform an unsupervised clustering of climate data from all different global climate models and classify the United States into a set of representative climate regions to immediately access the climate data for any target location. Climate data clustering quantifies uncertain projections of  precipitation and evapotranspiration (ET), thereby providing climate inputs for groundwater systems without downloading large amount of climate data.

%\input{Narrative/02models-surrogate}
%\label{sec:ufno}
\section{ML-based surrogate modeling for groundwater systems}\label{sec:emulator}

The intersection of physics and ML provides a rich space to build models with the advantages of high-dimensional data-driven learning while maintaining (and even guiding) physical constraints and laws. One popular method is called neural operator learning \citep{lu2019deeponet, li2020fourier, lavin21}, using neural network to learn mesh-independent, resolution-invariant solution operators for PDEs. Using neural operator learning, we aim to learn a fast surrogate model for groundwater systems with the physical simulation datasets from the Amanzi model. The input of our datasets includes uncertain soil, subsurface, and \textbf{climate properties (precipitation and ET: see Section~\ref{sec:climatepipeline})} $m(\textbf{x},t)$ for a groundwater system, the output includes the flow and transport properties $y(\textbf{x},t)$ such as the spatiotemporal contaminant concentration. $\textbf{x}$ is the location vector in 2D spatial cross-section, $t$ is the time variable ranging from 1954 to 2100. 

\subsection{Architectures: U-FNO-3D and U-FNO-2D}
We build two different neural operator architectures (U-FNO-3D and U-FNO-2D) to predict spatial-temporal contaminant concentration and groundwater flow properties. Both of these two architectures have Fourier Neural Operator (FNO) with enhanced U-Net architecture (U-FNO) \citep{li2020fourier,wen2022u}. FNO has linear transformations in Fourier frequency modes so that FNO can represent any resolution with additional back transformation. U-FNO adds an additional U-Net architecture for each Fourier layer, achieving a lower test error for multi-phase flow predictions \citep{wen2022u}. 
\begin{wrapfigure}[14]{r}{0pt}
  \centering
    \includegraphics[width = 0.7\textwidth]{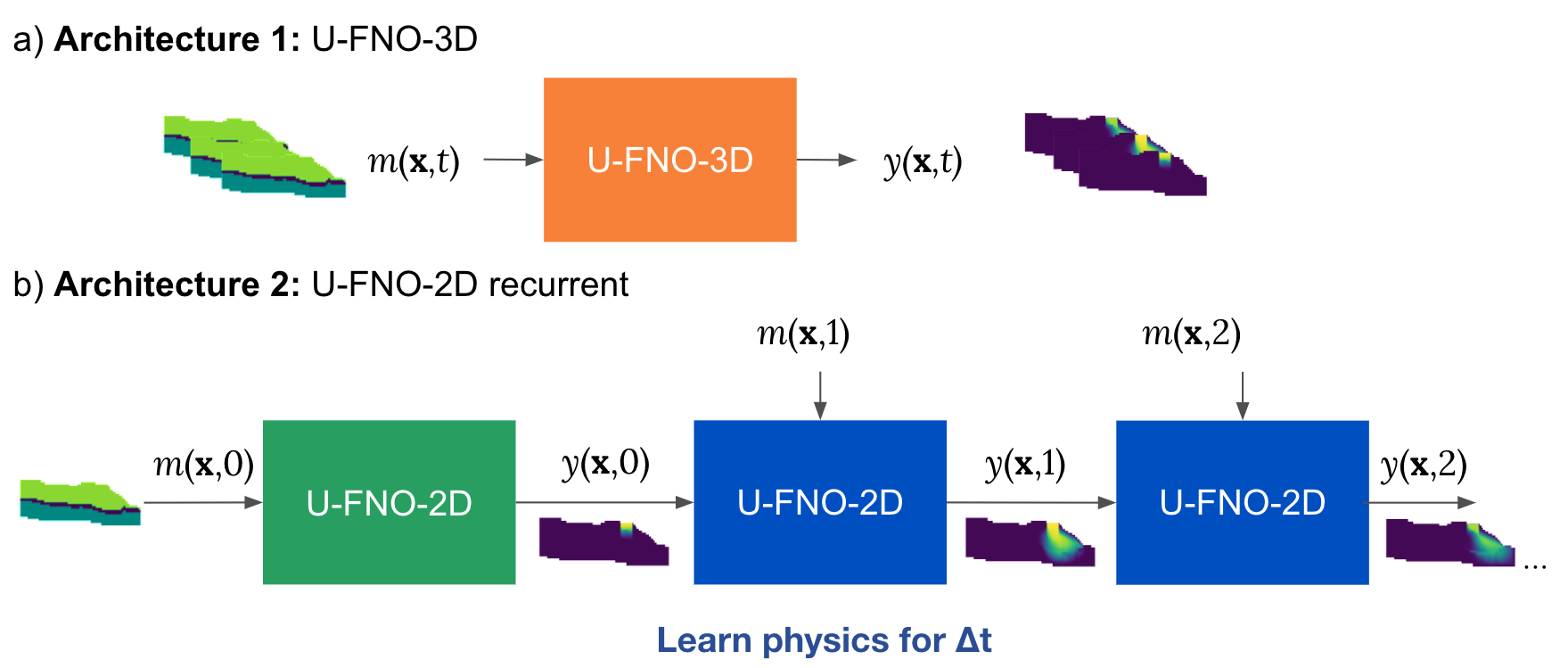}
    \caption{U-FNOs architectures}
    \label{fig:two_arch}
\end{wrapfigure}

Our two architectures address the temporal dimension differently.
\textbf{U-FNO-3D} takes all input time-series groundwater model parameters $m(\textbf{x},t)$ and predicts all contaminant concentrations for different time steps together $y(\textbf{x},t)$. 
\textbf{U-FNO-2D} uses a recurrent network: 
%the input and output for each time $t^*$ become $\{m(\textbf{x},t^*), y(\textbf{x},t^* - \Delta t)\}$ and $y(\textbf{x},t^*)$. 
U-FNO-2D learns the physics between every fixed time interval $\Delta t$. The initial stage $t = 0$ for U-FNO-2D does not have the prediction from the previous step. Then we train another U-FNO-2D model (the green block in Figure~\ref{fig:two_arch}b) for the initial stage $t = 0$ with the input $m(\textbf{x},0)$ and the output $y(\textbf{x},0)$. 
The advantage of U-FNO-2D is that we can preserve the time dependency, where the output $y(\textbf{x},t^*)$ is only determined by the input before and at time $t^*$.
However, this additional time dependency makes the training of U-FNO-2D practically harder and hence the training takes longer time. 
There are also accumulated errors through the recurrent neural network.

\subsection{Hybrid physics-based and data-driven loss functions}
We introduce four different loss functions that include both data-driven factors and physical boundary constraints. Our surrogate model predicts transient flow $\hat{h}(\textbf{x},t)$ and transport $\hat{c}(\textbf{x},t)$ properties $\hat{y}(\textbf{x},t) = \{\hat{h}(\textbf{x},t), \hat{c}(\textbf{x},t)\}$ where the ground truth are from numerical solvers. Therefore the designed loss functions target the data-driven mismatch between predictions $\hat{y} (\textbf{x},t)$ and $y(\textbf{x},t)$, and more interestingly, the physical constraints for solving PDEs such as boundary conditions. 

\textbf{Mean Relative Error}: We first quantify the data-driven mismatch using the mean relative error~(MRE) between $\ell_2$ norm:
\begin{equation}\label{eq:mre}
    \mathcal{L}_{MRE}(y,\hat{y}) = \frac{\|y-\hat{y}\|_2}{\|y\|_2} .
\end{equation}
\textbf{Spatial derivatives}: Additional mismatch on first derivatives in the horizontal direction $x$ and the vertical direction $z$ are also included.
\begin{equation}\label{eq:der}
    \mathcal{L}_{der}(y,\hat{y}) =  \frac{\|\partial{y}/\partial{x}-\partial{\hat{y}}/\partial{x}\|_2}{\|\partial{y}/\partial{x}\|_2}+\frac{\|\partial{y}/\partial{z}-\partial{\hat{y}}/\partial{z}\|_2}{\|\partial{y}/\partial{z}\|_2} .
\end{equation}
\textbf{Spatial derivatives on the contaminant boundary}: Maximum Contaminant Level (MCL) is the highest level of a contaminant that is allowed in drinking water recommended by the Environmental Protection Agency (EPA) \citep{libera2019climate}. Therefore, predicting the boundary of contaminant with higher concentration than the MCL is essential for site managers to protect water supply. We add first derivatives on the contaminant boundary $c \geq MCL$.

\begin{equation}
\begin{aligned}[b]
\mathcal{L}_{conc}(y,\hat{y}) = \frac{\|\partial{c'}/\partial{x}-\partial{\hat{c'}}/\partial{x}\|_2}{\|\partial{c'}/\partial{x}\|_2}+\frac{\|\partial{c'}/\partial{z}-\partial{\hat{c'}}/\partial{z}\|_2}{\|\partial{c'}/\partial{z}\|_2}, \\
\text{where } c'=
\begin{cases}
    0, & c < MCL \\
   1, & c \geq MCL \\
  \end{cases},\ \ \ \ \hat{c'}=\begin{cases}
    0, & \hat{c} < MCL \\
   1, & \hat{c} \geq MCL \\
  \end{cases}
  \end{aligned}
\end{equation}

\textbf{Physics-informed boundary conditions}: We add no flow boundary condition constraints in loss functions using physics-informed neural networks \citep{raissi2019physics} to help solving the PDEs. The boundary of the spatial domain $D$ is $\partial D$. 
\begin{equation}
\mathcal{L}_{BC}(\hat{y}) = \| \hat{h}|_{\partial D}\|_2 .
\end{equation}

The final loss function combines all above loss functions. 
\begin{equation}
\mathcal{L}(y,\hat{y}) = \mathcal{L}_{MRE}(y,\hat{y})+\beta_1 \mathcal{L}_{der}(y,\hat{y})+\beta_2\mathcal{L}_{conc}(y,\hat{y})+\beta_3\mathcal{L}_{BC}(\hat{y})
\label{eqn:loss_final}
\end{equation}

\section{Unsupervised climate data clustering}\label{sec:climatepipeline}
Climate data clustering leverages convolutional autoencoders to reduce the dimensionality of continental-scale climate simulation outputs and performs a spatial-temporal cluster analysis, which help stakeholders to get a realistic recharge rate immediately in order to evaluate the groundwater flow anywhere across the continental US~(CONUS). Note that median of precipitation and ET values from each resulting cluster will be used as inputs of the surrogate models (see in Section~\ref{sec:emulator}) in future work. 

We use monthly precipitation and ET values from \gfdl~\cite{dunne2012gfdl, dunne2013gfdl} that participated CMIP5 over CONUS. 
We remove the 3-month-running mean and then standardize the monthly data so that our deseasonalizing approach removes the recurrent patterns but remains the climatological anomalies.
We spatially subdivide each three months image into a 16 pixels $\times$ 16 pixels scale, $\approx$ $2^\circ$ $\times$ $2^\circ$ area, 
giving a smaller geographical and temporal unit, \emph{patch}. 
We train our approach with 70\% of only historical~(1950--2020) patches (we leave 30\% for inference) for 400 epochs on 4 K80 NVIDIA GPUs at a GCP instance to account for the future climate impact in inference.

\begin{figure*}[]
	\centering
	\begin{minipage}{.33\columnwidth}
		\centering
		\subfloat[Climate autoencoders.\label{fig:ae}]{
		    \includegraphics[width=1\columnwidth]{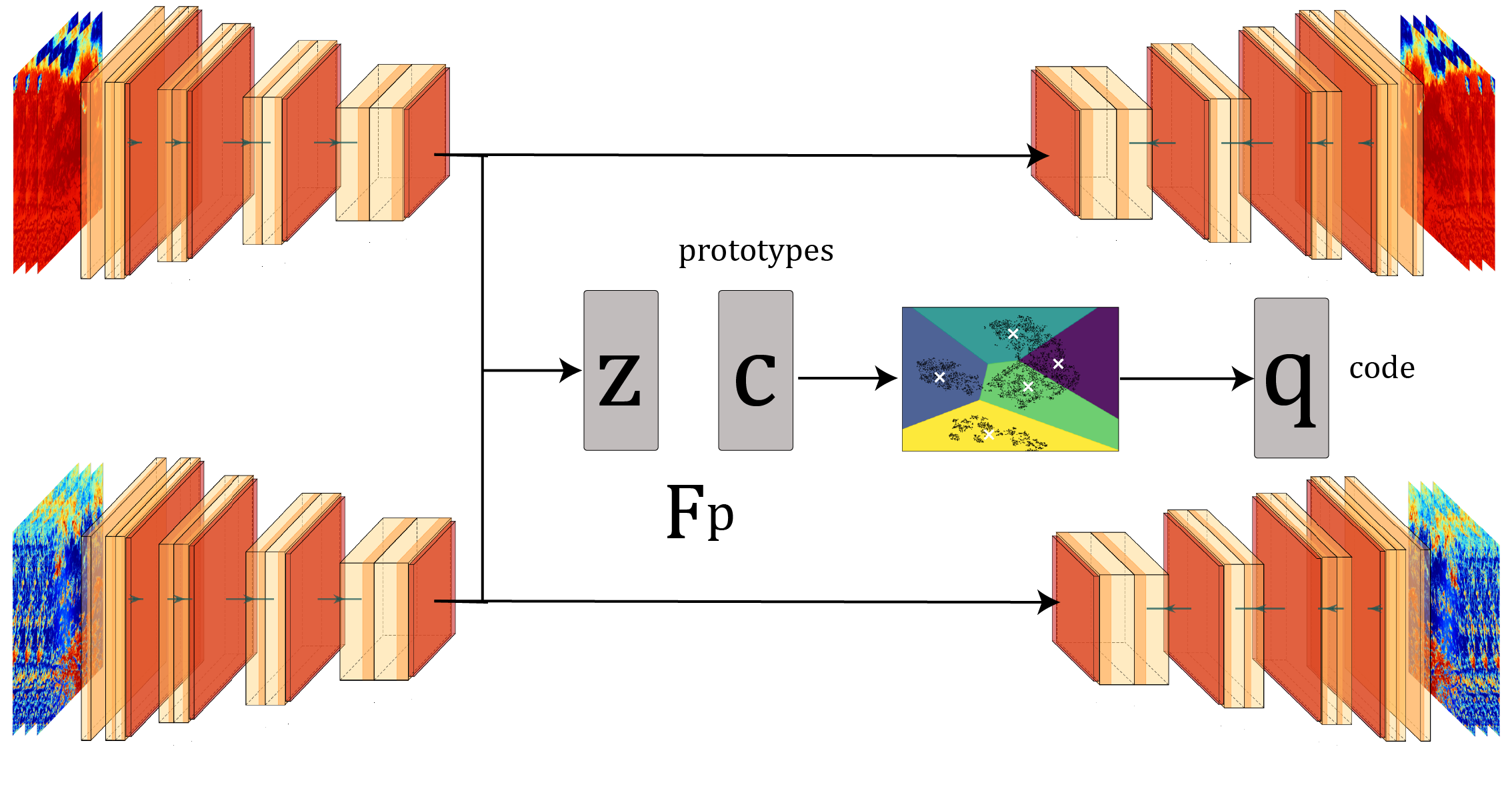}
		}\\
		\centering
		\subfloat[Clustering similarity.\label{fig:nmi}]{
		    \includegraphics[width=0.9\columnwidth]{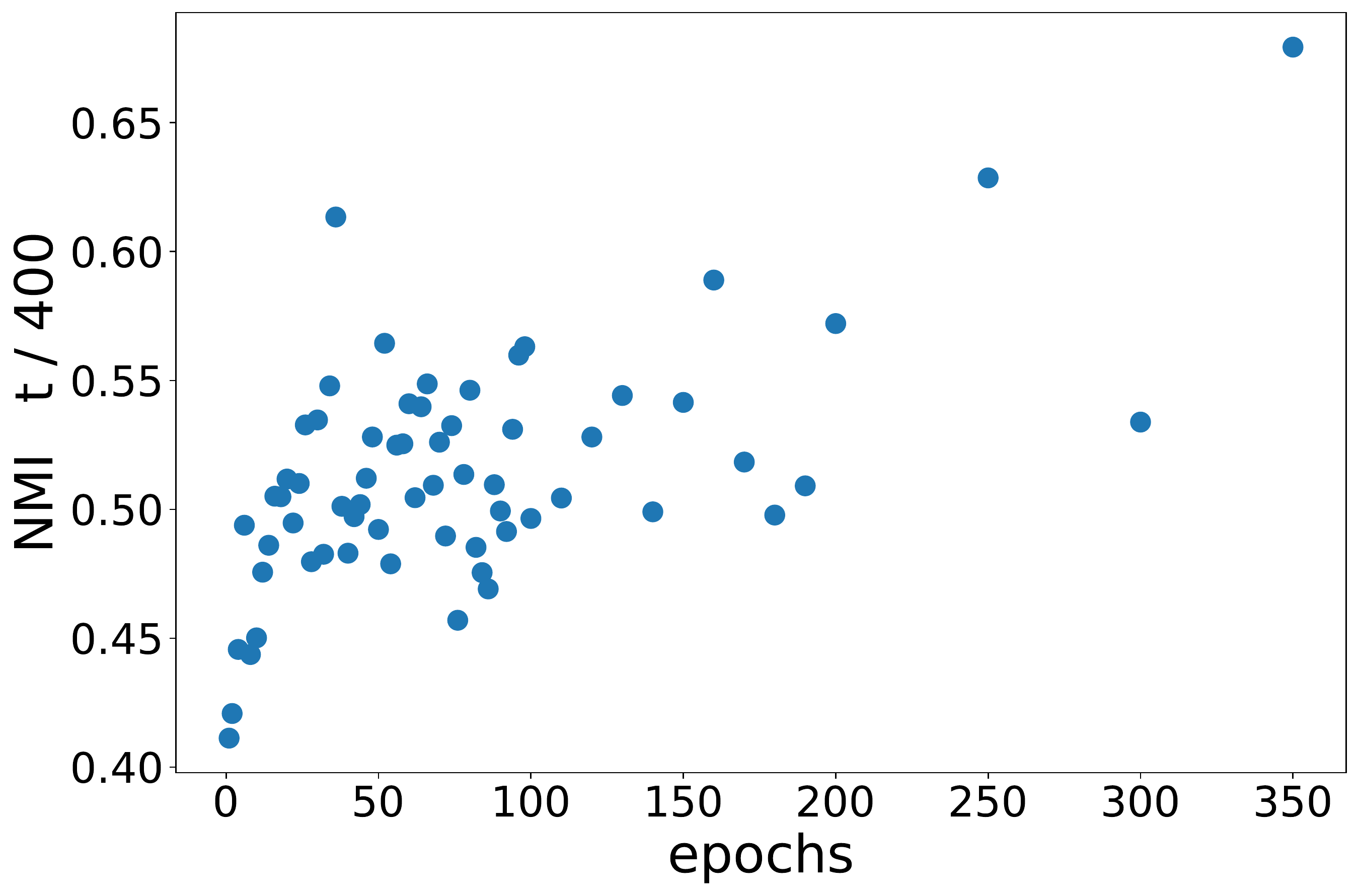}
		}
	\end{minipage}
	\begin{minipage}{.65\columnwidth}
		\centering
		\subfloat[Frequent cluster labels between 2061--2099.\label{fig:cluster}]{
		    \includegraphics[width=1\columnwidth]{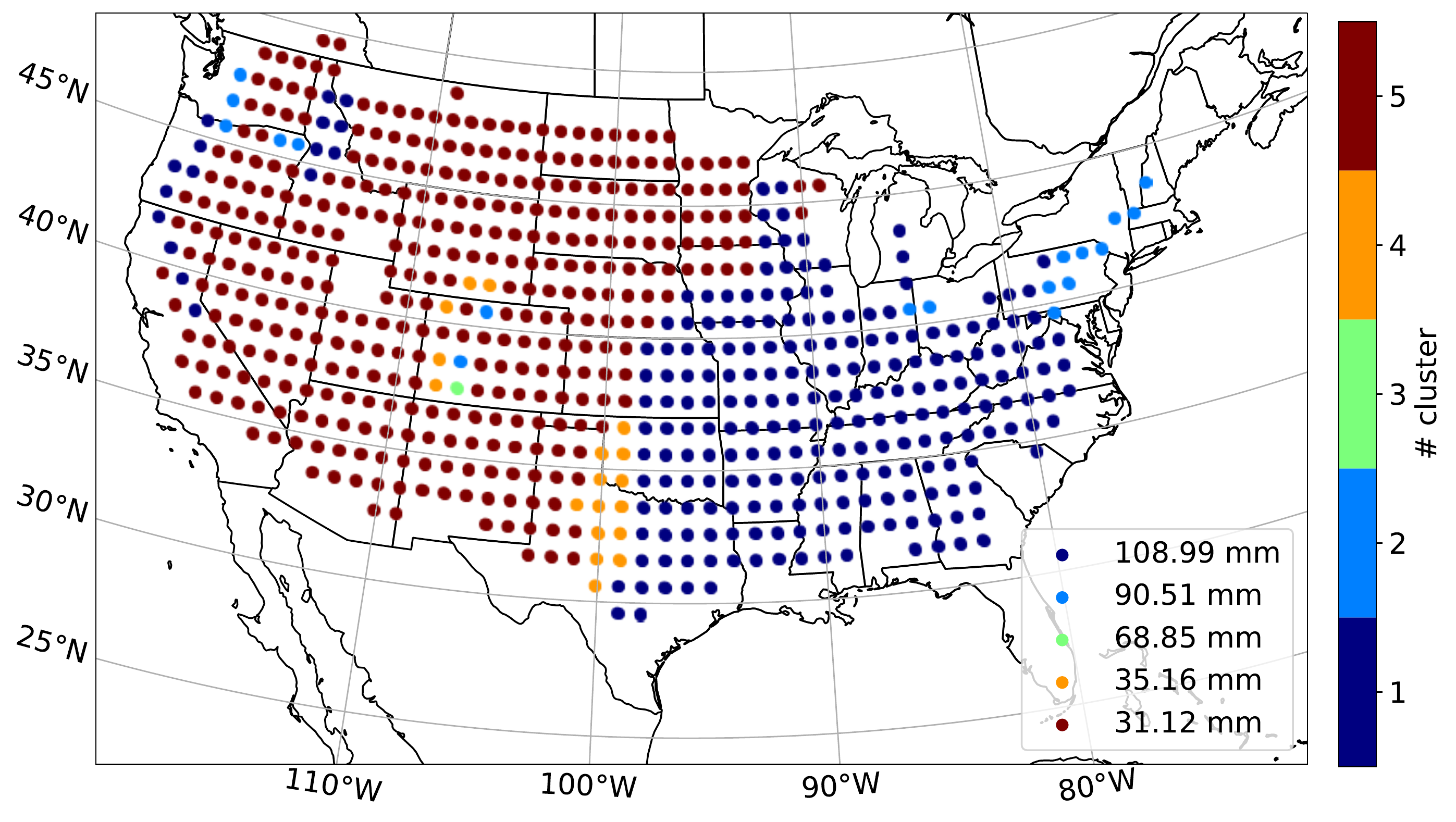}
		}
	\end{minipage}
	\caption{Architecture of climate autoencoders and evaluation of the results. (a) Features from two different autoencoders (i.e., precip. and ET.) are used for online clustering. (b) Similarity of cluster assignments along training. We calculate normalized mutual information (NMI) against labels from 400 epoch. Along training, clustering assignments evolves along training. (c) Spatial pattern of the most frequent cluster per grid between 2061--2099. Note that, we sort the clusters based on the mean precipitation within the cluster in descending order. Thus, small cluster numbers (e.g., \#1) represent wet climate.
	While the frequent cluster patterns dominant by \#1 and \#5, monthly cluster patterns show spatiotemporal diversity. 
	}
\end{figure*}

While scalable clustering algorithms are widely available~\cite{bahmani2012scalable}, an end-to-end autoencoder (see Figure~\ref{fig:ae}) and clustering training can further benefit the scalability.
A joint loss function~\cite{aparana2022clusteringreview} formulates a combination of both reconstruction and clustering loss terms:
\begin{equation}\label{eq:jointloss}
     \mathcal{L}_{\textrm{joint}}(\theta) = \lambda_{\textrm{reconst}}\mathcal{L}_{\textrm{reconst}}(\theta)   + \lambda_{\textrm{clustering}}\mathcal{L}_{\textrm{clustering}}(\theta)  , 
\end{equation}
where $\mathcal{L}_{\textrm{reconst}}$ corresponds to $L_2$ loss that quantifies the difference between a training data $x$ and output of autoencoder $D_{\theta}(E_{\theta}(x))$;  $\mathcal{L}_{\textrm{clustering}}$ term learns clustering assignments via online fashion; two coefficients $\lambda_{\textrm{reconst}}$ and $\lambda_{\textrm{clustering}}$ balance the two terms to achieve smoother optimization and better representations.
The learned representation via the end-to-end clustering approach reflects the association of group features of clusters by optimizing the joint loss function with Eq.~\ref{eq:jointloss}.

Our $\mathcal{L}_{\textrm{clustering}}$ is motivated by an online method~\cite{caron2020unsupervised} with simplifying the cross entropy loss between two cluster assignment: ``codes'' $q_c$ via the Sinkhorn-Knopp algorithm~\citep{cuturi2013sinkhorn} and $p_c$ that is computed as ``prediction'' obtained with a softmax of the dot product of K trainable prototype $\{c_1, \cdots, c_K \}$ and the latent representation $z_c = E_{\theta}(x)$. 
We calculate the dot product with an output layer from a single layer perceptron $F_p$ such that $ z_c^{\top}c = F_p(E_{\theta}(x))$. 
Thus the second loss term in Eq.~\ref{eq:jointloss} is 
\begin{equation}\label{eq:crossentropy_climate}
    \mathcal{L}_{\textrm{clustering}}(\theta) = - \sum_{k \in K} q_c^{(k)} \log{(p_c)} \quad \text{where} \quad  p_c^{(k)} = \frac{\exp{ \left(\frac{1}{\tau} z_c^{\top}c_k\right) }}{\sum_{k^{\prime}}\exp{\left(\frac{1}{\tau} z_c^{\top}c_k^{\prime} \right)}}.    
\end{equation}
Figure~\ref{fig:cluster} shows a spatial distribution of the most frequent cluster at each patch location for 2061--2099. Cluster numbers are sorted in descending order by the mean precipitation per patch.
Our cluster's climatological field is dominant by cluster \#1 (i.e., the wettest cluster) and \#5 (i.e., the driest climate cluster) over the CONUS.
We also observe that \#1 overlaps with humid subtropical climate zone (Cfa) and humid continental climate zone (Dfa) of a new and improved Koppen-Geiger climate classification 
%areas classified into C (temperate) and D (continental) types in K\"oppen–Geiger climate classification 
~\cite{beck2018present} and \#5 is spatially associated with their semi-arid (BS) and desert (BW) climate types, suggesting that our clusters capture physically meaningful spatial patterns.
%are changed frequently.

%\input{Narrative/04results}\label{sec:result}
\section{Evaluation of different architectures and loss functions}
We have in total 664 physical simulations from the Amanzi model with uncertain soil, subsurface and climate inputs. We split our dataset into training, validation, and testing (8:1:1) subsets. Table \ref{table:evaluation} (Index 1-4) shows the additional U-Net architecture gives us lower MRE and MSE on the validation dataset for both FNO-2D and FNO-3D architectures. In practice, U-FNO-3D trains around 4$\times$ faster (30 epochs, U-FNO-3D: 72 minutes, U-FNO-2D: 273 minutes). Therefore, we test our hybrid physics-based and data-driven loss functions on U-FNO-3D (Index 5-8 in Table \ref{table:evaluation}) . Every part of loss functions in Eq. \ref{eqn:loss_final}, when $\beta_i \neq 0$, reduces the MRE and MSE error. We have the lowest validation error when adding all spatial derivatives and no flow boundaries ($\beta_1 = \beta_2 = \beta_3 = 0.1$, Index 9-10, in Table~\ref{table:evaluation}). U-FNO-3D with all three combined loss functions achieves the lowest MRE and MSE after training 150 epochs on a A-100 NVIDIA GPU at a GCP instance).

\begin{table}[]
\vspace{-2ex}
\caption{Mean relative errors (MRE) and mean squared errors (MSE) for all experiments.}\label{table:evaluation}
\centering
\begin{small}
\begin{tabular}{lllllllp{1.5cm}}
\hline
Index & Architectures & Epochs & Loss $(\beta_1,\beta_2,\beta_3)$   & MRE & MSE & & Dataset\\ \hline 
1     & FNO-2D        & 30     & (0,0,0)       & 0.051                                                                            & 2.71e-4   & \rdelim\}{8}{0.1cm} &  \multirow{8}{1cm}{Validation} \\
2     & FNO-3D        & 30     & (0,0,0)       & 0.055                                                                            & 2.98e-4     & &                                                                     \\
3     & U-FNO-2D      & 30     & (0,0,0)       & \textbf{0.035}                                                                   & \textbf{1.51e-4}   & &                                                              \\
4     & U-FNO-3D      & 30     & (0,0,0)       & \textbf{0.037}                                                                   & \textbf{1.29e-4}  & &                                                               \\  \cline{1-6}
5     & U-FNO-3D      & 30     & (0.1,0,0)     & 0.029                                                                            & 8.83e-5           & &                                                               \\
6     & U-FNO-3D      & 30     & (0,0.1,0)     & 0.033                                                                            & 1.10e-4       & &                                                                   \\
7     & U-FNO-3D      & 30     & (0,0,0.1)     & 0.034                                                                            & 1.27e-4          & &                                                                \\
8     & U-FNO-3D      & 30     & (0.1,0.1,0.1) & \textbf{0.028}                                                                   & \textbf{8.14e-5}   & &                                                              \\ \cline{1-6}
9     & U-FNO-2D      & 150    & (0.1,0.1,0.1) & 0.020                                                                            & 4.49e-5 &     \rdelim\}{1.82}{*} & \multirow{2}{1cm}{Test}                                                               \\
10    & U-FNO-3D      & 150    & (0.1,0.1,0.1) & \textbf{0.014}                                                                   & \textbf{2.44e-5}  & &  \\\hline
\end{tabular}
\end{small}
\end{table}

\section{Conclusion}

In summary, we have successfully developed the ML-based multi-scale digital twin. Our ML-based surrogate model predicts spatiotemporal flow and transport properties immediately, saves computational times on solving PDEs, and thereby supports rapid decision-making for site managers. 
Our proposed unsupervised approach reduces the dimensionality of the vast historical and projected climate data to capture five unique climate patterns and provides lightweight climate properties for surrogate modeling. 
We believe that more climate resilience analysis for other contamination sites can benefit from our method in this paper to develop the groundwater flow and contaminant transport surrogate model with climate uncertainty. 

\section*{Broad Impact}
With tools for fast and reliable contaminant plume prediction under future climate scenarios, site managers and decision makers can evaluate the potential consequences and take rapid actions. We believe that more climate resilience analysis for other contamination sites can benefit from the method utilized in this paper to develop the groundwater flow and transport surrogate model. The use of clustering in the latent space of autoencoders on climate data for building representative climate regions can be extended to other applications in addition to using AI to create surrogate models.
%\input{Narrative/Impact}\label{sec:impact}

%\newpage
\begin{ack}

This work has been enabled by the Frontier Development Lab Program (FDL) USA.  FDL is a collaboration between SETI Institute and Trillium Technologies Inc., in partnership with (Google Cloud, NVIDIA).
%(list all government agencies, and industrial/compute partners that contributed).
This material is based upon work supported by the Department of Energy [National Nuclear Security Administration] under Award Number DE-AI0000001.
  	Disclaimer:  This report was prepared as an account of work sponsored by an agency of the United States Government.  Neither the US Government nor any agency thereof, nor any of their employees makes any warranty, express or implied, or assumes any legal liability or responsibility for the accuracy, completeness or usefulness of any information, apparatus product or process disclosed, or represents that its use would not infringe privately owned rights.  Reference herein to any specific commercial product, process, or service by trade name trademark, manufacturer, or otherwise does not necessarily constitute or imply its endorsement, recommendation, or favoring by the US Government or any agency thereof.  The views and opinions of authors expressed herein do not necessarily state or reflect those of the US Government or any agency thereof.

%This work has been enabled by the Frontier Development Lab (FDL.ai). FDL USA is a collaboration between several government agencies, Department of Energy (DOE), National Aeronautics and Space Administration (NASA), and U.S. Geological Survey (USGS), SETI Institute, and Trillium Technologies Inc., in partnership with private industry and academia. This public/private partnership ensures that the latest tools and techniques in Artificial Intelligence (AI) and Machine Learning (ML) are applied to basic research priorities in support of science and exploration of material concerns to human kind.
\end{ack}

%\section*{References}
%\reftitle{References}
\bibliography{references}

%%%%%%%%%%%%%%%%%%%%%%%%%%%%%%%%%%%%%%%%%%%%%%%%%%%%
\newpage
\section*{Check list: }
 If you are using existing assets (e.g., code, data, models) or curating/releasing new assets...

\begin{enumerate}

\item For all authors...
\begin{enumerate}
  \item Do the main claims made in the abstract and introduction accurately reflect the paper's contributions and scope? \yes
    %\answerTODO{}
  \item Did you describe the limitations of your work? \yes
    %\answerTODO{}
  \item Did you discuss any potential negative societal impacts of your work? \na
    %\answerTODO{}
  \item Have you read the ethics review guidelines and ensured that your paper conforms to them? \yes
    %\answerTODO{}
\end{enumerate}

\item If you are including theoretical results...
\begin{enumerate}
  \item Did you state the full set of assumptions of all theoretical results? \na
    %\answerTODO{}
  \item Did you include complete proofs of all theoretical results? \na
    %\answerTODO{}
\end{enumerate}

\item If you ran experiments...
\begin{enumerate}
    \item Did you include the code, data, and instructions needed to reproduce the main experimental results (either in the supplemental material or as a URL)? \no~Codes and simulation data are proprietary but we are planning to make them publicly available. 
    %\answerTODO{}
    \item Did you specify all the training details (e.g., data splits, hyperparameters, how they were chosen)? \yes
    %\answerTODO{}
    \item Did you report error bars (e.g., with respect to the random seed after running experiments multiple times)? \no
    %\answerTODO{}
    \item Did you include the total amount of compute and the type of resources used (e.g., type of GPUs, internal cluster, or cloud provider)? \yes~See Section~4
    %\answerTODO{}
\end{enumerate}

\item If you are using existing assets (e.g., code, data, models) or curating/releasing new assets...
\begin{enumerate}
  \item If your work uses existing assets, did you cite the creators? \yes
    %\answerTODO{}
  \item Did you mention the license of the assets? \yes
    %\answerTODO{}
  \item Did you include any new assets either in the supplemental material or as a URL? \no
    %\answerTODO{}
  \item Did you discuss whether and how consent was obtained from people whose data you're using/curating? \na
    %\answerTODO{}
  \item Did you discuss whether the data you are using/curating contains personally identifiable information or offensive content? \na
    %\answerTODO{}
\end{enumerate}

\item If you used crowdsourcing or conducted research with human subjects...
\begin{enumerate}
  \item Did you include the full text of instructions given to participants and screenshots, if applicable? \na
    %\answerTODO{}
  \item Did you describe any potential participant risks, with links to Institutional Review Board (IRB) approvals, if applicable? \na
    %\answerTODO{}
  \item Did you include the estimated hourly wage paid to participants and the total amount spent on participant compensation? \na
    %\answerTODO{}
\end{enumerate}

\end{enumerate}

%%%%%%%%%%%%%%%%%%%%%%%%%%%%%%%%%%%%%%%%%%%%%%%%%%%%%%%%%%%%

\bibliographystyle{abbrvnat}

\appendix

%\section{Appendix}

%Optionally include extra information (complete proofs, additional experiments and plots) in the appendix.
%This section will often be part of the supplemental material.

\end{document}